\documentstyle[12pt]{article}
\def\notp{p\kern-4.5pt\hbox{$/$} }

\def\bard{d\kern-2.5pt\raise 3pt\hbox{-}}
\newcommand{\comment}[1]{}% anything in the argument of a \comment is 
\newcommand{\lsm}{L$\sigma$M}
\begin{document}
\addtolength{\baselineskip}{1ex}
\begin{center}
 
{\Large \bf Double Counting Ambiguities}\\[.5cm]
{\Large \bf in the Linear $\sigma$ Model}\\[1cm]
{\bf  A. Bramon}\footnote{Grup de Fisica Te\'orica, Universitat Aut\'onoma de Barcelona,
08193 Bellaterra (Barcelona), Spain.}\\
{\bf  Riazuddin}\footnote{Department of Physics, King 
	Fahd University of Petroleum
and Minerals, Dhahran 31261, Saudi Arabia.}\\
{\bf  M. D. Scadron}\footnote{Physics Department, University of Arizona, Tucson, AZ 85721, USA.}\\
\vspace{.5cm}
Department of Physics,\\
International Centre for Theoretical Physics, Trieste Italy\\[.5cm]
\end{center}
 
\bigskip
\bigskip
\bigskip
\bigskip
\begin{abstract}
We study the dynamical consequences imposed on effective chiral 
field theories such as the quark-level  SU(2) linear
$\sigma$ model (L$\sigma$M) due to the fundamental constraints 
of massless Goldstone pions, the normalization of
the pion decay constant and form factor, and the pion 
charge radius.  We discuss quark-level double 
counting L$\sigma$M ambiguities in the context of the 
Salam-Weinberg $Z = 0$ compositeness condition.  Then SU(3) 
extensions to the kaon are briefly considered. 

\vspace{1.00 in}
\end{abstract}

\newpage
\section*{1.  Introduction}

Most physicists believe that the chiral theory of QCD is Nature's way of binding quarks
(and gluons) into observable hadrons.  Although there is no double-counting problem in this QCD
scheme, it is a nonlinear and nonperturbative theory which has not yet been solved in the
low energy region.  Consequently it is still reasonable to consider an effective field
theory---the quark-level linear $\sigma$ model (L$\sigma$M)---involving both quark and meson 
loops.  In fact, the particle data group has just reinstated the nonstrange $\sigma$ in the 1996 tables [1].%don't mess with this one
 Such a L$\sigma$M might suggest double counting problems, because
at first glance it is not clear when a (bound state) $\overline q q$ pion
or sigma meson should be treated as an SU(2) L$\sigma$M elementary particle.
In this paper
we attempt to resolve the L$\sigma$M double-counting ambiguities while focusing on the calculation of the 
(i) pion mass in the chiral limit (CL), (ii) pion decay constant and pion form-factor
normalizations, (iii) pion charge radius.

The first test of any chiral field theory is its ease in satisfying the Goldstone theorem
[2] in the chiral limit,
$$
m_\pi =0. 
\eqno(1)
$$
\noindent
A second test involves the double-counting issue for the pion decay constant and pion form factor
normalizations.  A third test concerns the pion charge radius, which is now measured respectively to
be [3,4]
$$
r_\pi = (0.66 \pm 0.02) \rm \ fm,  \ \ \ \ r_\pi = (0.63 \pm 0.01) \rm \ fm.
\eqno(2)
$$
The latter value matches perfectly with the highly successful phenomenology of vector meson 
dominance (VMD), which predicts [5]
$$
r^{VMD}_\pi = \sqrt{6}/m_\rho \approx 0.63 \rm \ fm.
\eqno(3)
$$

In fact the original chiral field theories of the 1960's, the SU(2) 
nucleon-level L$\sigma$M [6]
and the Nambu-Jona-Lasinio four-fermion model (NJL [7]) recovered $m_\pi = 0$ in (1).  Indeed,
earlier Nambu [8] dynamically invoked axial current conservation when $m_\pi = 0$ to induce the
extremely useful notion of partial conservation of axial currents (PCAC) for $m_\pi \neq 0$, which assumes a more quantitative form for the L$\sigma$M[6].
But it took until 1979 for physicists to obtain the {\em quark}-level L$\sigma$M chiral-limiting
prediction [9]
$$
r^{L \sigma M}_\pi = \sqrt{N_c}/2\pi f_{\pi,CL} \approx 0.60 \ \rm \ fm,
\eqno(4)
$$
for $N_c=3$ and the chiral-limiting pion decay constant [10] $f_{\pi,CL} \approx 90$ MeV.

In Sec.~2 we first study the SU(2) quark-level L$\sigma$M in one-loop order and remind the reader how the null
pion mass Goldstone theorem (1) is satisfied.  Next we introduce the external axial-vector and vector
(photon) fields and demonstrate how the pion decay constant  and pion form factor are
self-consistently determined for quark loops (only) up to cutoff
$\Lambda \approx 750$ MeV.  Then we allow for the smaller meson ($\vec \pi$ and $\sigma$) one-loop order
graphs and show that the cutoff $\Lambda^\prime \approx 660$ MeV is reduced to the value where the 
$\sigma (650)$ (see eq.\ (17)) is almost no longer an elementary field but instead is a $q\overline{q}$ bound state.~%
This speaks directly to the problem of double-counting the $\vec \pi$ and
$\sigma$ particles in the L$\sigma$M context of $Z = 0$ compositeness
conditions.

In Sec.\ 3 we begin by showing that the quark loops (only) pion charge radius (for $\Lambda \approx
750$ MeV $> m_\sigma (650)$) indeed recovers eq.(4).  However adding in the much smaller meson loops
(for $\Lambda^\prime \approx 660$ MeV $\sim m_\sigma (650)$) instead leads to an infrared singularity, so 
apparently the $\sigma (650)$ has become a $q\overline{q}$ bound state when computing $r^{L\sigma M}_\pi$.  At this
point we dynamically generate the entire L$\sigma$M beginning from a simpler chiral quark model theory.
This naturally links $r^{L\sigma M}_\pi$ in eq.(4) with $r^{VMD}_\pi$ in eq.(3).

We extend this quark-level L$\sigma$M theory to SU(3) states in Sec.\ 4, and show how the null kaon
Goldstone theorem operates in one (quark)-loop order.  Also the connection between $r^{VMD}_K$
and $r^{L\sigma M}_K$ continues to hold as for the pion charge radius above.  
We draw our conclusions in Sec.\ 5.  In particular we suggest that the
quark-level L$\sigma$M with these double-counting issues resolved in the
context of the $Z = 0$ compositeness condition has experimental relevance
as a broad $\sigma$ meson of mass below 1 GeV.  Such a $\sigma$ meson has
in fact been detected in recent data analyses [1].

\section*{2.  Quark-level SU(2) linear sigma model}

Shifting to the true (chiral) vacuum with expectation values $\langle \sigma \rangle = 
\langle \vec\pi \rangle = 0$, the interacting part of the SU(2) quark-level L$\sigma$M
lagrangian density is 
$$
{\cal L}^{L\sigma M}_{int} = g' \sigma(\sigma^2 + \vec\pi^2) - (\lambda/4) (\sigma^2 +\vec\pi^2)^2
+ g \bar\psi (\sigma + i\gamma_5\vec\tau \cdot \vec\pi) \psi ,
\eqno(5)
$$
with meson-quark and meson-meson chiral couplings for quark mass $m_q$ (and again $f_\pi \approx
93$ MeV),
$$
g=m_q/f_\pi \ \ ,\ \ \  \ \ g'=(m^2_\sigma - m^2_\pi) /2 f_\pi = \lambda f_\pi.
\eqno(6)
$$
While (5) and (6) are indeed the original tree level results of refs.[6] (but now with quarks
replacing nucleons), the Goldstone pion $m_\pi = 0$ should be reviewed in a dynamical context as
determined by the lagrangian (5).

At tree level the axial currents are conserved $\partial \vec{A} = 0$ because the axial current divergence term
$\partial_\mu \gamma^\mu \gamma_5/2$ interferes destructively against the massless pion pole 
$g f_\pi \gamma_5$ due to the quark-level Goldberger-Treiman relation (GTR)
$$
m_q = f_\pi g.
\eqno(7)
$$
This dynamical approach of Nambu [8] has a tree-level L$\sigma$M Gell-Mann-Levy version
[6] due to eqs.\ (6).

In one-loop-order, the quark-level L$\sigma$M obeys the Goldstone theorem in an interesting manner.
Specifically, the quark bubble and tadpole graphs contributing to the pion
mass are depicted in Fig.1a, 
while the corresponding 
$\pi$ and
$\sigma$ meson bubble, quartic loops and tadpole graphs are depicted in Fig.1b.  In the chiral limit
(CL), the quark bubble plus tadpole loops (ql) for $u$ and $d$ quark flavors occurring in $N_c$ colors
contribute to the pion mass (squared) according to 
$$
m^2_{\pi,ql} = i8N_c g \left( -g + {2g'm_q \over m^2_\sigma} \right)
\int {\bard^4 p \over p^2-m^2_q} ,
\eqno(8a)
$$
with $\bard^4 p = (2\pi)^{-4} d^4 p$.  
Regardless of the overall quadratically divergent integral in (8a),  the quark loop component of 
$m^2_\pi$ vanishes if $g'$ is dynamically fixed to $g' = m^2_\sigma / 2 f_\pi$ in (6) when 
$m^2_\pi = 0$ because of the GTR eq.(7).

To handle the meson loops (ml) depicted in Fig.1b, one must first invoke the partial fraction identity
$$
{ g'^2 \over (p^2-m^2_\sigma) ( p^2-m^2_\pi) }= {\lambda  \over 2} \left[
{ 1 \over p^2-m^2_\sigma} - { 1 \over p^2-m^2_\pi} \right] , 
$$
with coupling coefficients again related using eqs.(6).  Then the quadratically divergent $\pi$
 and $\sigma$ parts separate into two vanishing integrals
$$
m^2_{\pi,ml} = (-2\lambda + 5\lambda - 3\lambda)i \int {\bard^4 p \over p^2 - m^2_\pi}
+ (2\lambda + \lambda - 3\lambda)i \int {\bard^4 p \over p^2 - m^2_\sigma}.
\eqno(8b)
$$
Again the coupling coefficients multiplying these two divergent integrals in (8b) identically
vanish.  Therefore the complete one-loop level Goldstone theorem in the L$\sigma$M
becomes in the chiral limit [11]
$$
m^2_\pi = 0|_{quark\ loops} + 0|_{\pi\ loops} + 0|_{\sigma\ loops} = 0.
\eqno(9)
$$

While the L$\sigma$M Goldstone theorem (9) is not surprising, the vanishing of (9) places no
constraints on $m_\sigma$ or on the cutoff involved.  As we shall see, however, the analogue 
L$\sigma$M one-loop versions of the pion form factor normalization and the pion charge radius
will put severe constraints on the (ultraviolet) cutoff.   This will in turn instruct us when the
scalar $\sigma$ particle can indeed be taken as an elementary particle, or should instead be 
treated as a $\bar q q $ bound state.

But first we consider the pion decay constant $f_\pi = m_q /g$ in (7) to one-loop order in the 
L$\sigma$M as given by the quark loop (ql) and meson loop (ml) diagrams depicted in Fig.2a,b 
respectively.  The dominant $u$ and $d$ quark loops of Fig.2a generate a fermion trace
4$m_qq_\mu$ in the chiral limit, so the (constituent) quark mass factor cancels out, leading
to the ``log-divergent gap'' equation
$$
1 = -i 4 N_c g^2 \int\limits^{\Lambda} \bard^4 p (p^2 - m^2_q)^{-2} = \ln (X + 1) - [1+X^{-1} ]^{-1}.
\eqno(10a)
$$
Here $X = \Lambda^2 / m^2_q$, so for $N_c = 3$ and $g \approx 3.6$ (as we shall later show
but estimate here from the GTR $g \approx 320$ MeV/90 MeV $\sim 3.6$), the numerical solution
of (10a) is $X \approx 5.3$ or $\Lambda \approx \sqrt{5.3} \ \ m_q \approx 750$ MeV
for $m_q \approx 325$ MeV (as we shall later find).  This self-consistent cutoff separates the
elementary $\sigma$ meson $< 750$ MeV from the $\bar qq$ bound states: $\rho(770), 
\omega(783), A_1(1260) > \Lambda \approx 750$ MeV.

If we add to the quark loops of Fig.2a the (smaller) meson loop of Fig.2b, the cutoff $\Lambda$
in (10a) remains essentially unchanged.  More specifically, the pion to vacuum matrix element
of the axial-vector current $if_{\pi} q _\mu $ is incrementally shifted to  
$$
i \delta f_\pi q_\mu = 2g' \int {\bard^4 l (2l+q)_\mu \over 
(l^2 - m^2_\sigma) [(l+q)^2 - m^2_\pi] } .
\eqno(10b)
$$
Changing Feynman variables to $l \to l - qx$ in (10b), and accounting for the (linearly
divergent) surface term [12], the net dimensionless
shift of $f_\pi$ due to the meson loop of Fig.2b is 
$$
\delta f_\pi / f_\pi = \lambda / 16\pi^2  - i 2 \lambda \int^1_0 dx(1-2x)
\int {\bard^4 l \over [l^2 - m^2_\sigma (1-x)]^2} .
\eqno(10c)
$$

However explicit evaluation of the latter (log-divergent) integral in (10c) when folded into the
vanishing integral $\int^1_0 dx(1-2x)$ in fact leads to the finite contribution $- \lambda / 16\pi^2$,
which {\em precisely} cancels the surface term $+ \lambda / 16\pi^2$.   Thus one is led back
to the log-divergent gap equation (10a) {\em even after} meson (as well as quark) loops are
included in Figs.2.
This minimal shift in the PCAC relation is of course expected, and in fact has
already been used when analyzing the $\pi^+ \to e^+ \nu \gamma$ form factors in a quark-level
L$\sigma$M context [13].

This log-divergent gap equation (10a)  is in fact recovered from the quark loop
version of the pion form factor normalization $F_\pi (q^2 = 0) = 1$.  Then the quark loops
(ql) probed by the off-shell (vector) photon in Fig.3a lead to the pion form factor in
the chiral limit [14]
$$
F^{ql}_\pi (q^2) = -i 4 N_c g^2 \int^1_0  dx \int \bard^4 p [p^2 - m^2_q + x (1-x)q^2]^{-2}.
\eqno(11a)
$$
The apparent linear divergence of $F^{ql}_\pi (q^2)$ is removed in (11a) by rerouting one-half
of the loop momentum in the opposite direction [14], with the pion form factor covariant defined as
$<\pi^+ | V_\mu | \pi^+ > = F_\pi (q^2)(p'+p)_\mu$.  Then at $q^2=0$, the quark loop pion form
factor in (11a) is normalized by (10a):
$$
F^{ql}_\pi (q^2 = 0) = -i 4 N_c g^2 \int^\Lambda \bard^4 p [p^2 - m^2_q]^{-2}  = 1,
\eqno(11b)
$$
{\em provided} the cutoff is $\Lambda \approx 750$ MeV as found from the pion decay constant
combined with the quark-level GTR in eqs.(10).

However this satisfying result (11b) is {\em significantly altered} when the L$\sigma$M
meson loops (ml) are included in the pion form factor as depicted in Fig.3b.  While the
second (quartic) pion loop vanishes, the first $\pi^+ \sigma$ loop in Fig.3b contributes
to the pion form factor covariant as
$$
F^{ml}_\pi(q^2)(p'+p)_\mu = (2g')^2 i \int {\bard^4 l (2l+p'+p)_\mu \over
(l^2 - m^2_\sigma)[(l+p')^2 - m^2_\pi] [(l+p)^2 - m^2_\pi]}.
\eqno(12a)
$$
Since the meson loop integral in (12a) is convergent, one can shift Feynman variables
to $l \to l + (qx - p')y $ and pick out the $(p'+ p)_\mu$ covariant from (12a) to obtain the 
meson loop component of the pion form factor,
$$
F^{ml}_\pi(q^2) = 2 m^2_\sigma \lambda i \int^1_0 dx \int^1_0 dy 2y (1-y) \int
\bard^4 l [l^2 + q^2 x (1-x)y^2 - m^2_\sigma (1-y)]^{-3},
\eqno(12b)
$$
in the chiral limit.  Here we have used $(2g')^2 = 2 m^2_\sigma \lambda$ from eq.(6).  Finally invoking
the dynamically generated meson coupling strength [15] $\lambda = 2g^2 = 8\pi^2/3$ (to
which we shall later return in Sec.\ 3), one obtains the overall pion form factor normalization
adding (11a) to (12b) at $q^2 = 0$ and using the cutoff $\Lambda'$ version of (10a):
$$
1 = F^{ql}_\pi(0) + F^{ml}_\pi(0) = \ln ( X' +1 ) - [ 1 + X'^{-1}]^{-1} + {1 \over 6}.
\eqno(12c)
$$
Here $X'={\Lambda'}^2/m^2_q$ and the $m^2_\sigma$ mass term in (12b) cancels out [12], resulting in the additional
ml factor $\lambda/16\pi^2 = 1/6$ in (12c).

The numerical solution of (12c) is $X' \approx 4.15$ or $\Lambda' \approx \sqrt{4.15}$ $m_q 
\approx 660$ MeV.  This latter cutoff scale is {\em dangerously close} to the NJL $\sigma$ mass
scale $m_\sigma = 2 m_q \approx 650$ MeV (this latter $\sigma$ mass also holds in the 
dynamically generated version of the L$\sigma$M [15]).  It suggests that for the pion form 
factor, the ``elementary" $\sigma$ meson at $\sigma (650)$ may be double-counting its
$\bar qq$ bound state version.  This cutoff problem becomes a major issue when the derivative
of the pion form factor is taken as needed for computing the pion charge radius in Sec.\ 3 to follow.

The above form factor normalization problem based on the log-divergent gap equation (10a) or (11b) vs.\ (12c) is an example of the Salam-Weinberg [16] $Z = 0$ compositeness condition.
This condition gives a self-consistent field-theoretic 
interpretation of a $\overline{q}q$ pion and sigma meson treated {\em either} as elementary particles (as in the L$\sigma$M) or as bound states (as in the NJL model).  In this case the resulting 
inequalities $m_\sigma \approx 650$ MeV $\approx \Lambda' \approx 660$ MeV $< \Lambda \approx 750$ MeV $< m_\omega \approx 780$ MeV speak to the Salam-Weinberg compositeness condition.

\section*{3.  Pion charge radius in L$\sigma$M and VMD theories}

Having reconfirmed the exact pion form factor normalization due to the dominant quark loops and
also the approximate normalization for the meson loop corrections of order 15\%,  we now focus 
on the  pion charge radius $r_\pi$, where differentiation of $F_\pi (q^2)$ at $q^2 = 0$ will 
lead to no inconsistencies for pure quark loops (ql) since then $\Lambda \approx 750$ MeV.  However,
when meson loops are included, $F_\pi (0)=1$ only when $\Lambda \to \Lambda' \approx $ 660 MeV, which
means $\sigma(650)$ is on the verge of becoming a $\overline{q}q$ bound state.

Specifically for quark loops (ql) alone and $\Lambda \approx 750$ MeV, the pion charge radius in
the chiral limit is 
$$
r^2_{\pi , ql} = {6 d F_{\pi,ql} (q^2) \over d q^2} |_{q^2=0} 
 = {-i4N_c g^2 (-2) \over (2\pi)^4} \int^1_0 dx 6 x (1-x) \int {d^4p \over 
(p^2 - m^2_q)^3 }
$$
$$
 = { N_c \over 4\pi^2 f^{2}_{\pi,CL}}, \ \ \  
\eqno(13a)
$$
upon using the GTR $g^2/m^2_q = 1/f_{\pi, CL}^2 $.  This of course is the result of refs.[9], which
can also be obtained via a once-subtracted dispersion relation evaluated at $q^2 = 0$:
$$
r^2_{\pi,ql} = {6\over\pi} \int^\infty_0 dq^2 {Im F_\pi (q^2) \over (q^2)^2} = {N_c \over 
4\pi^2 f_{\pi,CL}^2}.
\eqno(13b)
$$
We stress again the {\em uniqueness} and {\em finiteness} of $r_\pi$ in eqs. (13) for quark loops 
and for the VMD value of $r_\pi$ in (3).   Since the (quark model) cutoff $\Lambda \approx 750$ MeV in
(10a) has shifted to the lower value of $\Lambda' \approx 660$ MeV in (12c), it is legitimate to consider 
only quark loops even in the quark-level L$\sigma$M when computing $r_\pi$ via differentiation of
$F_\pi (q^2)$ as in (13).  In effect, the shift of $\Lambda > m_\sigma$ to $\Lambda' \approx m_\sigma$
means that the elementary $\sigma (650)$ meson in the L$\sigma$M is becoming a $\bar qq$ bound state as
in the NJL scheme.  As such, the L$\sigma$M picture with $m_\sigma = 2 m_q$ is merging into a pure quark
model or NJL picture, again with $m_\sigma = 2 m_q$.

Next we turn to $r_\pi$ as obtained from the meson loops (ml) of Fig.3b and eq.(12b).  
Differentiating (12b) with respect to $q^2$ and afterwards setting $q^2=0$, one obtains the 
meson loop (ml) contribution to the pion charge radius in the chiral limit,
$$
r^2_{\pi,ml} = {6(-2m^2_\sigma \lambda) \over (2\pi)^4 }i \int^1_0 dx x (1-x) \int^1_0
dy 2y (1-y) y^2 \int {d^4 l \over [l^2 - m^2_\sigma  (1-y)]^4 }
$$
$$
= {1\over36}  {1\over m^2_\sigma} \int^1_0 dy 2y^3 (1-y)^{-1},
\eqno(14)
$$
where the latter Feynman integral in (14) is $i \pi^2 [6m^4_\sigma (1-y)^2]^{-1}$, and again
$\lambda = 8\pi^2/3$.  Although the squared length scale in (14) is 150 times smaller than the VMD
scale $r^2_\pi = 6/m^2_\rho$, the infrared singularity in (14) is signaling that the derivative of
the meson loop form factor $F^{ml}_\pi (q^2)$ (with normalization cutoff $\Lambda' \approx 660$ MeV
$\approx  m_\sigma$) has finally led to an inconsistency because this $\sigma (650)$ must then be
treated as a ${\overline q}q$ bound state when computing $r^{L\sigma M}_\pi$.  
Stated another way, the formal infrared singularity of $r_{\pi , ml}$  in (14) characterizes chiral
symmetry {\em breakdown} since then the log divergence in (14) corresponds to $\ln m_\pi/m_\sigma \to
\infty$ as $m_\pi \to 0$.  This is a second signal (along with (12c)) that meson loops in a L$\sigma$M
may lead to a double-counting inconsistency (resolved by a $Z = 0$ compositeness condition).  This justifies the 
pure quark loop 
treatment of $r_\pi$ in eqs.(13a) and (13b), and it should not be surprising that the
$r^{L\sigma M}_\pi$ is then in close agreement with experiment.

As for the relation between the one-loop order L$\sigma$M approach and tree-level VMD, 
with quark loops alone for $\Lambda \approx 750$ MeV $< m_\rho$, the
rho meson can be taken as an external (bound state) particle and then the log-divergent gap equation in
eq.(10a) leads to [15,17]
$$
g_{\rho\pi\pi} = g_\rho [-i4N_c g^2 \int^\Lambda \bard^4 p (p^2-m^2_q)^{-2} ] = g_\rho.
\eqno(15a)
$$
This  is Sakurai's [5] VMD universality condition.  Equation (15a) can also be interpreted as a Z=0 
compositeness condition [16] for the L$\sigma$M.  
If meson loops such as in Fig.3b (with $\gamma\to\rho^\circ$) are included in eqs.(12), the extension of 
(15a) is  
$$
g_{\rho\pi\pi} = g_\rho [-i4N_c g^2 \int^\Lambda \bard^4 p (p^2-m^2_q)^{-2} ]  + {1\over6}  g_{\rho\pi\pi} 
\eqno(15b)
$$
in a dynamically generated L$\sigma$M context.  Here [15] $\lambda / 16\pi^2 = \frac {1}{6}$ as in eq.(12c).  However
since the (external field) rho meson is still a $\bar qq$ bound state, we still maintain that the (quark model)
cutoff is $\Lambda \approx 750$ MeV as in (10a) or (10c).  Then eq.(15b) 
becomes
$$
g_{\rho\pi\pi} = g_\rho + {1\over6}  g_{\rho\pi\pi} \ \ or \ \ g_{\rho\pi\pi} / g_\rho 
= {6\over 5} \approx 1.2,
\eqno(15c)
$$
and the latter ratio is in good agreement with data:  $g_{\rho\pi\pi} \approx 
6.1$ follows from the $\rho$ width
and $g_\rho\approx 5.1$ follows from the $\rho^\circ \to e^+ e^-$ decay rate.

Since $r^{VMD}_\pi \approx 0.63$ fm in eq.(3) (for $\rho$ as an external field with $\Lambda < m_\rho$)
and $r^{L\sigma M}_\pi \approx 0.60$\,fm in eq.\ (4) ~(then for quark loops
alone with UV cutoff\newpage
\noindent
$\Lambda \approx 750$ MeV $< m_\rho$), there may be even a deeper link between 
$r^{VMD}_\pi$ and $r^{L\sigma M}_\pi$ in the chiral limit.  We now probe for such a connection.

To study the Goldstone theorem for $m^2_\pi$ and also the pion form factor $F_\pi (q^2 = 0)$ in Sec.\ 2
and the pion charge radius $r_\pi$ in Sec.\ 3, we have used only the original L$\sigma$M lagrangian
in eqs.(5,6) (but for quark rather than for nucleon fields). We have alluded to the dynamically generated
L$\sigma$M [15] only to streamline the results.  Besides eqs.(5,6), the dynamically generated L$\sigma$M
appeals to dimensional regularization to obtain the two additional relations [15]
$$
m_\sigma = 2m_q \ \ \ , \ \ \ g = 2\pi/\sqrt{N_c} \approx 3.6276
\eqno(16)
$$
for $N_c = 3$.  Of course the former equation in (16) is the famous NJL relation [7], while the latter
together with the GTR (7) predicts a sensible chiral-limiting nonstrange quark mass
$$
m_q = f_{\pi,CL} 2\pi/\sqrt{3} \approx (90 \rm MeV) (3.6276) \approx 325 MeV,
\eqno{(17a)}
$$
and in turn a scalar sigma mass 
$$
  m_\sigma = 2 m_q \approx 650 \mbox{MeV}.
\eqno{(17b)}
$$

	The log-divergent gap equation (10a) helps to dynamically generate the rho couplings in (15) to one
loop-order by invoking the VMD version of the $\rho$ to vacuum matrix element of the em vector 
current $<0|V^{em}_\mu | \rho^o > = ek^2\varepsilon_\mu ) / g_\rho$ 
for $k^2 = m^2_\rho$.  
Then the quark loop for the latter $\gamma-\rho^o$ transition in the soft limit leads to [15]
$$
g_{\rho\pi\pi} = g_\rho = \sqrt{3} g = 2\pi ,
\eqno(18)
$$
a result also obtained by other methods [18].  Note that (18) is numerically compatible with
the measured $\rho\pi\pi$ coupling constant extracted from the $\rho$ width, giving 
$g^2_{\rho\pi\pi} /4\pi \approx 3.0$ or $|g_{\rho\pi\pi}| \approx 6.1$.

Furthermore recall Sakurai's derivation [19] of the KSRF relation from VMD of the $I=1$ $\pi N \to
\pi N$ scattering amplitude: $M^{(-)} = g^2_\rho / m^2_\rho$.  This is to be equated with the current
algebra form $M^{(-)} = 1/2f^2_\pi$, leading to the KSRF relation [20] 
(ignoring the slight 15\% correction from (15))
$$
m^2_\rho = 2 f^2_\pi g^2_\rho,
\eqno(19a)
$$
which is empirically accurate to 10\%.  This is justified for $\rho \to \pi\pi$ because
momentum consrvation requires $p_\rho \to 0$ when $p_\pi$ and 
$p_\pi' \to 0$.  Combining (19a) with the
dynamical generated L$\sigma$M scale (18) and the quark level GTR (7) then converts the KSRF relation
to
$$
m_\rho = \sqrt{2} f_\pi g_\rho = \sqrt{6} f_\pi g =  \sqrt{6}m_q \approx 795 \rm MeV.
\eqno(19b)
$$
Moreover using $m_\rho =\sqrt{6}m_q$ in (19b) then transforms the VMD relation for the pion charge
radius in (3) to
$$
r^{VMD}_\pi = \sqrt{6}/m_\rho = 1/m_q =   \sqrt{3}/2\pi f_{\pi,CL} = r^{L\sigma M}_\pi.
\eqno(20)
$$
Thus we have dynamically linked $r^{VMD}_\pi$ to $r^{L\sigma M}_\pi$, as anticipated.
 
Although $r^{VMD}_\pi$ in (3) and  $r^{L\sigma M}_\pi$ in (4) appear numerically close, the dynamically
generated versions of $r^{VMD}_\pi$ and $r^{L\sigma M}_\pi$ become {\em identical} in the chiral
limit (CL) due to the blending together of the L$\sigma$M with VMD via KSRF.  
The expression $r_\pi = (1/m_q)$ in (20) suggests a quark model interpretation
of the pion charge radius for a $\overline q q$ Goldstone pion.  Namely when
$m_\pi \rightarrow 0$ the quarks {\it fuse} together with (Coulombic-type)
potential $1/r$ and relativistic (Compton-type) wave length $r_\pi = (1/m_q) 
\approx 0.60$ fm, in close agreement with observation.   
As such, eq.(20)
places a tight VMD-KSRF-L$\sigma$M constraint on other models purporting to contain all the 
richness of chiral symmetry.

\section*{4.  Extension to SU(3) linear $\sigma$ model and VMD}

Here we show that the natural generalization of the SU(2) linear $\sigma$ model (L$\sigma$M) discussed
in Secs.\ 2 and 3 but now for
the SU(3) L$\sigma$M also driven by  the quark-level Goldberger-Treiman relation
(GTR) gives [9,21]
$$
f_\pi g = \hat m \ \  \ ,  \ \ \ f_K g = {1\over2} (m_s + \hat m ).
\eqno(21)
$$
Then the ratio of the two GTRs in (21) eliminates the meson-quark coupling $g$ and predicts the 
empirical ratio
$$
{f_K \over f_\pi} = {1\over2} ( {m_s \over \hat m} + 1) \approx 1.22 \ \  or \ \ 
{m_s \over \hat m} \approx 1.44
\eqno(22)
$$
Indeed, this latter strange to nonstrange constituent quark mass ratio is approximately  obtained from
baryon magnetic moments [22], meson charge radii [23] and from $K^* \to K \gamma $ decays [24].

Note that we have not passed to the SU(3) $\times$ SU(3) chiral limit in (21) or (22).  But
we do so now when studying the SU(3) generalization of the Goldstone theorem for $m^2_K = 0$
in a L$\sigma$M context.  Then the quark loops $(ql)$ in Fig.4 generate a straightforward extension
of the SU(2) result in eq.(8):
$$
m^2_{K,ql} = i 4N_c g \int \bard^4 p \left[ {-2 g (p^2 - m_s \hat m) \over (p^2 - m_s^2) (p^2 - \hat m^2)}
+ {2 g_{NS}' \hat m \over  m_{\sigma_{NS}}^2 (p^2 - \hat m^2)} 
+ {\sqrt{2} g_{S}' m_s \over  m_{\sigma_S}^2 (p^2 - m_s^2)} \right]  \ .
\eqno(23)
$$
Here $\sigma_{NS}$ represents the SU(2) nonstrange $\sigma$ meson and $\sigma_S$ is the SU(3)
$\bar ss$ extension.  That the integrand in (23) in fact {\em vanishes} can be seen from  the
partial fraction identity
$$
{(m_s + \hat m) (p^2 - m_s \hat m) \over (p^2 - m_s^2) (p^2 - \hat m^2)}=
{\hat m \over p^2 - \hat m^2} + { m_s \over p^2 - m^2_s} ,
\eqno(24)
$$
combined with the natural [25] SU(3) extensions of the L$\sigma$M meson-meson coupling in (6),
$g' = m^2_\sigma / 2 f_\pi$:
$$
g'_{NS} = {m^2_{\sigma_{NS}} \over 2 f_K }\ \ \ ,  \ \ \ 
g'_{S} = {m^2_{\sigma_S} \over \sqrt{2} f_K }.
\eqno(25)
$$
The same is true for the meson loop graphs for $m^2_K$.  Thus the SU(3) Goldstone theorem
$m^2_K = 0$ indeed holds in a straightforward manner to one-loop order in the SU(3) L$\sigma$M.
Note that as in the SU(2) L$\sigma$M version of the Goldstone theorem (8) and (9), the
vanishing of $m^2_K$ in (23) is independent of any precise values of the scalar mesons.

The SU(3) analysis for the kaon charge radius $r_{K^{+}}$ is even more transparent than the 
Goldstone vanishing of $m^2_K$ in (23) for the L$\sigma$M.  Since $r^{-1}_{\pi^+} = \hat m$ in
(20) naturally corresponds to the kaon extension $(m_s + \hat m ) /2$, use of the SU(3) GTRs
in (21) along with $g=2\pi/\sqrt{3}$ from (16) leads to the charge radius
$$
r^{L\sigma M }_{K^+} = \sqrt{3} / 2\pi f_K \approx 0.49 \ \ \rm fm 
\eqno(26)
$$
for $f_K \approx 110 $ MeV in the chiral limit.  On the other hand the SU(3) VMD extension of
eq.(3) is
$$
r^{VMD}_{K^+} = \sqrt{6} / m_{K^*} \approx 0.54 \ \ \rm fm .
\eqno(27)
$$
Not only are (26) and (27) in close proximity, but both are near the observed value  
$<r^2_{K^+}>  \approx 0.28  fm^2 $ or $r_{K^+} \approx 0.53 fm$.

Stated another way, the authors in ref.[23] develop a constituent quark mass expansion for the
($L\sigma$M) quark loop version of the $r_{\pi^+}$ and $r_{K^+}$ charge radii ratio:
$$
{ <r^2_{K^+}> \over <r^2_{\pi^+}>} \approx 1 - (5/6) \delta + (3/5) \delta^2 \approx 0.75 , 
\eqno(28)
$$
for $\delta = (m_s / \hat m) - 1 \approx 0.44$ from eq.(22).  This compares quite well with the 
measured ratio $0.70 \pm 0.12$.  The extension to the (neutral) kaon charge radius in ref.\ [23] is
also reasonable
$$
{ <r^2_{K^0}> \over <r^2_{\pi^+}>} \approx (-1/3) \delta + (1/2) \delta^2 \approx -0.05 , 
\eqno(29)
$$
whereas data finds the latter ratio to be $-0.12 \pm 0.06$.

\section*{5.  Summary}

We have attempted to resolve the apparent ambiguities arising in a 
quark-level linear sigma model (L$\sigma$M) field theory to one-loop order.
In Sec.\ 2 we have shown that both quark and meson loops in the SU(2) 
L$\sigma$M manifest the Goldstone
theorem and the pion decay constant combined with the quark-level GTR.  While the same is true for quark
loops generating the pion form factor (and the log-divergent gap equation for cutoff $m_\sigma <
\Lambda \approx 750$ MeV $< m_\rho$), adding in meson loops to $F_\pi (q^2)$ reduces the cutoff to 
$\Lambda' \approx 660$ MeV $\approx  m_\sigma$.  This suggests that the $\bar qq$ $\sigma$ is shifting from
an elementary L$\sigma$M particle to a NJL bound state when one extracts
 $F_\pi(q^2)$ at $q^2 = 0$.
 
We began Sec.\ 3 by computing the quark loop version of the pion charge radius $r_\pi$, and the
result is of course finite and in good agreement with experiment.  However when L$\sigma$M meson
loops are included, $r_\pi$ develops an infrared singularity.  This just means that the $\sigma$
meson must be taken as a ${\overline q}q$ bound state (since the pion form factor cutoff is $\Lambda' \approx 660$
MeV $\approx m_\sigma$) when computing the pion charge radius.  Then we invoked the dynamically generated 
L$\sigma$M theory finding $g_{\rho\pi\pi} = 2\pi$ (also compatible with data) together with the chiral KSRF
relation $m_\rho = \sqrt{2} f_\pi g_\rho$ and showed that in the CL, $r^{L\sigma M}_{\pi , ql}
= r^{VMD}_\pi$ exactly.

The above calculations employed the SU(2) dynamically generated $L\sigma M$
requiring [15] $m_\sigma = 2 m_q,\quad g = 2 \pi / \sqrt 3.$  Since the former
relation also follows from the four-fermion theory of NJL [7] where the 
$\vec \pi$ and $\sigma$ are $\overline q q$ bound states (so there is no
meson ambiguity), it should not be surprising that the quark-level $L\sigma M$
also has no $\vec \pi, \sigma$ elementary particle --bound state
ambiguity (due to the $Z = 0$ compositeness condition).

Finally in Sec.\ 4 we extended the L$\sigma$M to SU(3) and demonstrated that the kaon Goldstone theorem for
quark loops is again manifest.  We also computed the K$^+$ charge radius r$^{VMD}_{K^+}$ in tree order and r$^{L \sigma M}_{K ^+}$ in one-loop order.  Both are compatible with data.

With hindsight, double-counting problems never arise in QCD or in the NJL 
four-quark pictures because only quarks (and gluons in the former case) are
elementary while mesons are treated as $\overline q q$ bound states.  The
quark-level L$\sigma$M in one-loop order (but with the double-counting
issues discussed in this paper taken into consideration) has the additional
scales of $m_q \approx 325$ MeV and $m_\sigma = 2 m_q \approx 650$ MeV 
dynamically generated [15].  As pointed out in the latter reference, the
log-divergent gap equation for the pion decay constant in our (10a) (or the
pion form factor normalization in (11b)) can be taken as a $Z = 0$ 
compositeness condition [16] characterizing the $\pi$ and $\sigma$ particles
as not elementary, but bound states of more basic fields (as in QCD or in
NJL).  This $Z = 0$ compositeness condition ((10a), (11b) or (15a)) in turn
bootstraps quark loop graphs to tree diagrams.  Such a ``nonperturbative 
shrinkage'' justifies not adding contact terms to one-loop terms as one would
do in a (multiple-counting) perturbative field theory.

While focusing on this issue of double counting 
in the quark-level \lsm , we have also obtained new 
one-loop order results:  (1) the pion decay constant 
involving both quark and meson loops; (2) the 
normalization of the pion form 
factor $F_\pi (q^2 = 0) = 1$ involving 
both quark and meson loops; (3) recovering the VMD 
universality relation $g_{\rho \pi \pi} = g_\rho$ due to 
quark loops
only and extending it including also 
meson loops to the coupling ratio $g_{\rho \pi \pi} / g_\rho = 6/5$ 
which is in empirical agreement with the 
observed $\rho \pi \pi$ and $\rho \overline{e} e$ 
decay rates; (4) using the KSRF relation to link the \lsm\ 
pion charge radius $r_\pi$ to the VMD version 
of $r_\pi$; (5) empirically extending $r_\pi$ to $r_K$ in 
the \lsm\ and VMD models. One might objectively question why we have
worked so hard to repair this ``toy theory''---the quark-level 
linear $\sigma$ model---using the Salam-Weinberg $Z = 0$ 
compositeness condition.  Our answer is that it is now 
becoming experimentally clear in [1] 
and [26--30] that a broad $\sigma$ meson of 
mass below 1 GeV is emerging from data (just as the dynamically 
generated L$\sigma$M requires).

Specifically, the DM2 collaboration [26] obtained a 
low mass $\pi \pi$ scalar $M \approx 420$ MeV 
from $J/\psi \to \omega \pi \pi$, while a 
reanalysis [27] of CERN-Munich data 
for $\pi^- p \to \pi^- \pi^+ n$ found a $\sigma$ mass near 850 MeV.
More recently a T\"ornqvist and Roos 
data analysis [28] finds a very 
broad $\sigma$ meson at $f_0$(400--900), with an 860 
MeV mass coming from a Breit-Wigner background and 
its pole at 400--900 GeV.  Also Svec [29] 
studied polarized target $\pi N \to \pi 
\pi N$ data and detected a $\sigma$(750).  Finally, 
Ishida et.\ al.\ [30] analyzed the $\pi \pi$ scattering 
phase shifts and introduced a negative background phase 
and found a $\sigma$(555) scalar meson. Also  see ref. [31].

	We conclude that a chiral $\sigma$ meson 
may indeed exist and that the quark-level L$\sigma$M 
with a $\sigma$(650) may not be simply a ``toy'' model but in fact may reflect the real world.
\eject
{\bf  Acknowledgements}:  The first author (AB) receives partial financial support from Spanish Government grants DGICYT, AEN 93/520.
The latter two authors (R and MDS) appreciate discussions with N. Paver and partial support from
ICTP in Trieste.  Also R receives support from KFUPM, and MDS from
the Universitat
Aut\'onoma de Barcelona and INFN Italy.
\eject

\newpage

\leftline{\bf  Figure Captions}

\bigskip
\noindent
Fig.1a.  Quark bubble plus tadpole graphs for $m^2_\pi$.

\bigskip
\noindent
Fig.1b.  Meson bubble plus quartic loop plus tadpole graphs for $m^2_\pi$.

\bigskip
\noindent
Fig.2.  Quark (a) and meson (b) loops contributing to $f_\pi$.

\bigskip
\noindent
Fig.3a.  Quark loops contributing to the pion form factor $F^{ql}_\pi (q^2)$.

\bigskip
\noindent
Fig.3b.  Meson loops contributing to $F^{ml}_\pi (q^2)$.

\bigskip
\noindent
Fig.4.  Quark bubble plus tadpole graphs for $m^2_K$.

\end{document}